\documentclass[10pt,times,twocolumn]{article}
\usepackage{lineno,hyperref}
\modulolinenumbers[5]

\usepackage{amssymb}
\usepackage{amsthm}
\usepackage[affil-it]{authblk} % format author list
\usepackage[T1]{fontenc} % correct font encoding
\usepackage{amstext}
\usepackage{amsmath} % math package
\usepackage{multirow}
\usepackage{textcomp} % allows use of \textdegree to make degree symbol
\usepackage{tensor} % allows for sub/superscripts in math mode before a variable
\usepackage{indentfirst}
\usepackage{pdflscape}
\usepackage{multicol}
\usepackage{array}
\usepackage{shadow}
\usepackage{floatflt}
\usepackage{wrapfig}
\usepackage{picinpar}
\usepackage{framed}
\usepackage[free-standing-units=true]{siunitx}
\usepackage{longtable,lscape} % allows the use of multi-page tables, also sideways.
\usepackage{outlines} % for outlining things.  
\usepackage{times}
\usepackage{float}
\usepackage{fixltx2e} % allows use of \textsubscript{}... \textsuperscript{} already included
\usepackage{lipsum} % generates random text
\usepackage[normalem]{ulem} % strike through etc without emphasis change
\usepackage{graphicx}    % Add some packages for figures. Read epslatex.pdf on ctan.tug.org
\usepackage{color}
\usepackage{epsfig}
\usepackage{longtable,lscape} % allows the use of multi-page tables, also sideways.
\usepackage{rotating} % allows for 'sidewaystable'
\usepackage{hyperref}
\usepackage[left=0.5in, right=0.5in, top=1in, bottom=1in]{geometry}
\usepackage{setspace}
\usepackage{abstract}
\usepackage{tikz} % for drawing graphics 
\usepackage{xspace} % smart space 
\usepackage{xcolor}
\usepackage{soul} % text highlighting 
\usepackage[showdow, en-US]{datetime2}
\DTMsetstyle{mmddyyyy}
\usepackage{mathrsfs}

\usepackage[belowskip=5pt,aboveskip=5pt]{caption}
\usepackage{subcaption}
\usepackage{sectsty}
\sectionfont{\normalsize\normalfont\bfseries}
\subsectionfont{\normalsize\normalfont\itshape}
\subsubsectionfont{\normalsize\normalfont\itshape}
\makeatletter
\renewcommand{\@seccntformat}[1]{\csname the#1\endcsname.\quad}
\makeatother

\usepackage{upgreek}
\usepackage{chemmacros}
\usepackage{textgreek} 
\chemsetup{greek=textgreek}
\chemsetup{modules=all}
\chemsetup[reactions]{
	before-tag = R,
	tag-open = ( ,
	tag-close = )
	}
\chemsetup[chemformula]{}
\usepackage{tikz}
 
\newlength \figWidthCol
\newlength \figWidthFull
\newcommand{\app}{{\raise.17ex\hbox{$\scriptstyle\sim$}}} % tilde ~

\newcommand{\ie}{\textit{i.e.}} % id est --> that is --> In other words
\newcommand{\supinf}{Supplementary Information\xspace}
\newcommand{\fom}{\ensuremath{\mathscr{F}}}

\usepackage[numbers, square, comma, sort&compress]{natbib}
\begin{document}

\title{Relativistic light sails need to billow\\ }

\author[a]{Matthew F.\ Campbell}
\author[b]{John Brewer}
\author[c]{Deep Jariwala}
\author[b]{Aaswath Raman}
\author[a]{Igor Bargatin\thanks{Corresponding author. \\ Email: \href{mailto:bargatin@seas.upenn.edu}{bargatin@seas.upenn.edu}}}
\affil[a]{Department of Mechanical Engineering and Applied Mechanics, University of Pennsylvania, Philadelphia PA, USA 19104}
\affil[b]{Department of Materials Science, University of California at Los Angeles, Los Angeles CA, USA 90024}
\affil[c]{Department of Electrical and Systems Engineering, University of Pennsylvania, Philadelphia PA, USA 19104}

\date{\today}

%Articles start with a fully referenced summary paragraph, ideally of no more than 200 words, which is separate from the main text and avoids numbers, abbreviations, acronyms or measurements unless essential. It is aimed at readers outside the discipline. This summary paragraph should be structured as follows: 
%2-3 sentences of basic-level introduction to the field; 
%a brief account of the background and rationale of the work; 
%a statement of the main conclusions (introduced by the phrase 'Here we show' or its equivalent); 
%and finally, 2-3 sentences putting the main findings into general context so it is clear how the results described in the paper have moved the field forwards. 

% 

\twocolumn[
  \begin{@twocolumnfalse}
    \maketitle
	\begin{onecolabstract}
		{We argue that light sails that are rapidly accelerated  to relativistic velocities by lasers must be significantly curved in order to reduce their mechanical stresses and avoid tears. Using an integrated opto-thermo-mechanical model, we show that the diameter and radius of curvature of a circular light sail should be comparable in magnitude, both on the order of a few meters in optimal designs for gram-scale payloads. Moreover, when sufficient laser power is available, a sail's acceleration length decreases and its chip payload capacity increases as its curvature increases. Our findings provide guidance for emerging light sail design programs, which herald a new era of interstellar space exploration. \\ }
	\end{onecolabstract}
  \end{@twocolumnfalse}
]
\saythanks

Our knowledge about life on other planets is currently limited by the energy density of chemical and nuclear fuels~\cite{Cohen2019-37, OReilly2021-22} and by the resolution of Earth-based and orbiting telescopes~\cite{Wagner2021-922}. Light sails with nanometer-scale thicknesses and meter-scale diameters that carry gram-scale microchip payloads~\cite{Niccolai2019-1} and are accelerated to relativistic velocities by photons from high-power lasers have the potential to significantly increase humankind's understanding of the cosmos~\cite{Marx1966-22, Redding1967-588, Forward1984-187, Lubin2016-40, Parkin2018-370, Atwater2018-861}. Recent discoveries of exoplanets in the Alpha Centauri~A/B system~\cite{Dumusque2012-207, Anglada-Escude2016-437, Damasso2020-eaax7467, Wagner2021-922} have prompted the Starshot Breakthrough Initiative~\cite{Lubin2016-40, Parkin2018-370} to set Proxima Centauri, a star located about 4.2~light years away from Earth, as a target destination for light sail development. Thus far, studies have examined light sails in terms of their optical properties and radiative heat transfer~\cite{Ilic2018-5583, Myilswamy2020-8223, Salary2020-1900311, Weiliang2020-2350}, beam-riding stability~\cite{Popova2016-1346, Manchester2017-L20, Myilswamy2020-8223, Srivastava2019-3082, Siegel2019-2032, Gao2020-SF3J6, Srivastava2020-570}, and interaction with the interstellar medium~\cite{Early2015-205, Hoang2017-5, Ocker2021-inPress}. 

One important aspect that has received less attention, however, is the fact that the photon pressure required to accelerate a light sail will lead to significant stress and strain in the sail film. This could potentially dictate the survivability of the sail and ultimately determine the success of the entire mission. Several authors have examined the mechanics of \textit{solar} sails, which are sunlight-driven analogs to laser-propelled light sails. However, solar sails have order-of-magnitude larger diameters and generally experience much lower photon pressures and accelerations than photon-driven light sails, such that the primary concern of photon-induced deformation in solar sails is trajectory alteration rather than sail breakage~\cite{Zhang2020-2204, Huang2021-2613}. To our knowledge, only a single paper has considered structural mechanics issues for a laser-powered light sail undergoing high acceleration, in particular focusing on the wrinkling instabilities of a square planar sail and providing bending stiffness design guidelines~\cite{Savu2020-3842}. We are unaware of any work examining the light-pressure-induced stresses and strains in curved sails, which benefit from their passive laser beam-riding stability~\cite{Popova2016-1346}. 

\begin{figure}[t]
\centering
\includegraphics[width=\figWidthCol]{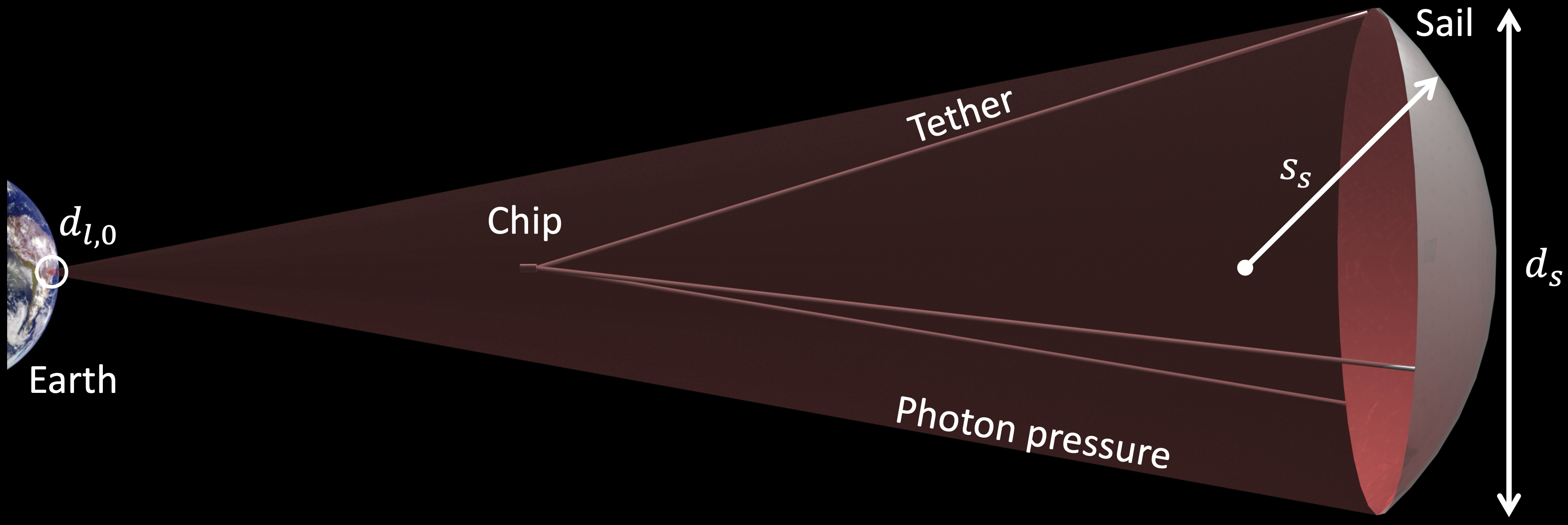}%
\caption{Concept diagram of a curved circular light sail towing a chip using three tethers. The sail has a perpendicular-to-beam diameter $d_s$ and spherical radius of curvature $s_s$, and is accelerated by an array of Earth-bound lasers whose diameter is $d_{l,E}$. Earth image obtained from NASA~\cite{NASA1998-photo}.}%
\label{F:overviewFigure}%
\end{figure}

Here, we therefore consider a generic spherically curved circular sail with gram-scale mass, square-meter-scale area, and submicron-scale thickness that tows a gram-scale-mass Starchip payload~\cite{Niccolai2019-1} (Figure~\ref{F:overviewFigure}), and analyze its photon-flux-induced stress and strain in terms of key design dimensions and material property limitations. Our analysis reveals that, because of mass constraints and radiative heating and cooling factors, practical sails of interest \textit{must} be significantly curved in order to survive the light pressure from the laser beam. This conclusion mirrors the behavior of parachutes and wind sails of sailboats, which can also minimize wind-induced stresses by billowing~\cite{Heinrich1966-52, Gordon1978-book}, as well as the typical designs of lightweight pressure vessels, which use curved (spherical or cylindrical) rather than flat walls. To identify optimal sail configurations, we also propose a new figure of merit that incorporates the acceleration length as well as thermal and mechanical factors, and use this metric to show that small increases in acceleration length allow greater thermal and stress failure margins and increase the likelihood of successful launches. Furthermore, we demonstrate that, given sufficient laser power, as the curvature of a sail increases, it is able to tow more massive chip payloads and simultaneously achieve lower acceleration distances. 

We begin by introducing three high-level light sail design trade-offs that will guide our numerical analysis. Consider the light sail pictured in Figure~\ref{F:overviewFigure}, which has diameter $d_s$, spherical radius of curvature $s_s$, and sail film thickness $t_f$. The first trade-off involves the sail curvature, which increases (the sail becomes more curved) as $s_s$ is reduced (at constant $d_s$). Beyond a certain point, specifically for $s_s<\frac{d_s}{\sqrt{2}}$, laser light incident on the perimeter of the sail will be reflected toward the sail's axial center, which constitutes a multiple-reflection process that is associated with additional absorption and heating. Conversely, if $s_s$ is made to be too large (low curvature), the resulting nearly flat sail may not only be unstable on the laser beam~\cite{Popova2016-1346} but also may experience excessive bending or tensile stresses, as we discuss in detail below. The second trade-off incorporates the sail film thickness $t_f$. Sails that are very thick tend to exhibit high absorption of incident laser light and thus overheat, whereas sails that are vanishingly thin exhibit low reflectivities and therefore accelerate slowly. The third trade-off concerns the diameter $d_s$. For a fixed incident laser power that is perfectly focused on the sail area, sails with smaller diameters experience greater photon intensities and are subject to overheating. In contrast, since the minimum sail film thickness has a lower bound of a few atoms at best, sails with large diameters must become more massive and hence will accelerate more slowly. Importantly, the previous point illustrates that these sail design parameters are interrelated. In this case, for a given sail mass $m_s$, $t_f$ scales inversely with the film density $\rho$ and sail surface area $A_s$, which in turn depends on $d_s$ and $s_s$: $t_f=\frac{m_s}{\rho A_s}$. These trade-offs demonstrate the significance and complexity of the variables $d_s$, $s_s$, and $t_f$ in light sail engineering. 

We now proceed to a more detailed treatment of light sail mechanics in order to understand the intricate interactions of these design parameters more fully. To accomplish this, we adopt a simplified sail model that uses a single effective material based on the properties of molybdenum disulphide (\ch{MoS2}) and alumina (\ch{Al2O3}), which have desirable optical, mechanical, and manufacturability attributes for light sails~\cite{Atwater2018-861}. As discussed in the \supinf, the single-layer effective-material model adequately describes the behavior of multilayer composite sails made from  \ch{MoS2} and \ch{Al2O3} while greatly simplifying the optical, thermal, and mechanical analyses. Despite the simplicity of this model, the fundamental insights it uncovers should translate to more complex sail designs and future alternative materials as well~\cite{Haastrup2018-042002, Ilic2020-769}. 

For simplicity, our effective material model adopts temperature-independent mechanical properties (density $\rho$, Young's modulus $E$, yield stress $\sigma_{yield}$, and critical strain $\epsilon_{crit}$) equivalent to those of thin-film polycrystalline \ch{MoS2} at 300~\si{\kelvin}~\cite{Ghobadi2017-1483, Graczykowski2017-7647, Sledzinska2020-1169}, and also uses the ultra-high vacuum sublimation temperature of bulk \ch{MoS2} $T_{sub}\approx1000$~\si{\kelvin}~\cite{Cannon1959-1612, Cui2018-44} as the thermal limit. For the effective optical properties,  we use a 90\%/10\% weighted average of the room-temperature complex index of refraction values ($n$ and $\kappa$) of \ch{MoS2}~\cite{Ermolaev2020-21} and \ch{Al2O3}~\cite{Lingart1982-706, Querry1985-report}, respectively. 

Using this model, we first examine the sail's acceleration length, photon pressure, and equilibrium temperature, and determine how these scale in relation to fundamental design properties. The acceleration length $L$ (see \supinf), or the distance the sail travels while being illuminated by laser-generated photons, is a key figure of merit for light sail design; it scales as $L\sim\frac{m_{tot} c v_f^2}{\Phi_0 \varrho_a}$, where $m_{tot}$ is the total mass of the sail, chip, and tethers, $c$ is the speed of light, $v_f$ is the final sail velocity, $\Phi_0$ is the laser output power, and $\varrho_a$ is the sail's average reflectivity near the laser wavelength. As the sail accelerates over this distance, it will feel an effective photon pressure $P$, caused by the impinging and reflecting laser light, which scales as $P\sim\frac{\Phi_0 \varrho_\perp}{c A_\perp}$, where $A_\perp$ is the perpendicular-to-laser beam area of the sail and $\varrho_\perp$ is the perpendicular-to-sail reflectivity at the sail's axial center (often, $\varrho_\perp\approx\frac{4}{3}\varrho_a$)~\cite{Parkin2018-370}. In addition, the sail will absorb a small fraction of the incoming laser power and achieve an equilibrium temperature $T$ at which it radiatively emits an equal amount of power. This temperature scales as $T^4\sim\frac{\alpha_a \Phi_0}{\varepsilon_e \sigma A_\perp}$, where $\alpha_a$ and $\varepsilon_e$ are the average absorptivity near the laser wavelength and the effective hemispherical emissivity of the sail, respectively, and $\sigma$ is the Stefan?Boltzmann constant. We note that, in general, the emissivity and absorptivity are themselves functions of temperature, but in this simplified analysis we have fixed them at their room temperature values.

\begin{figure}[t]
\centering
\includegraphics[width=\figWidthCol]{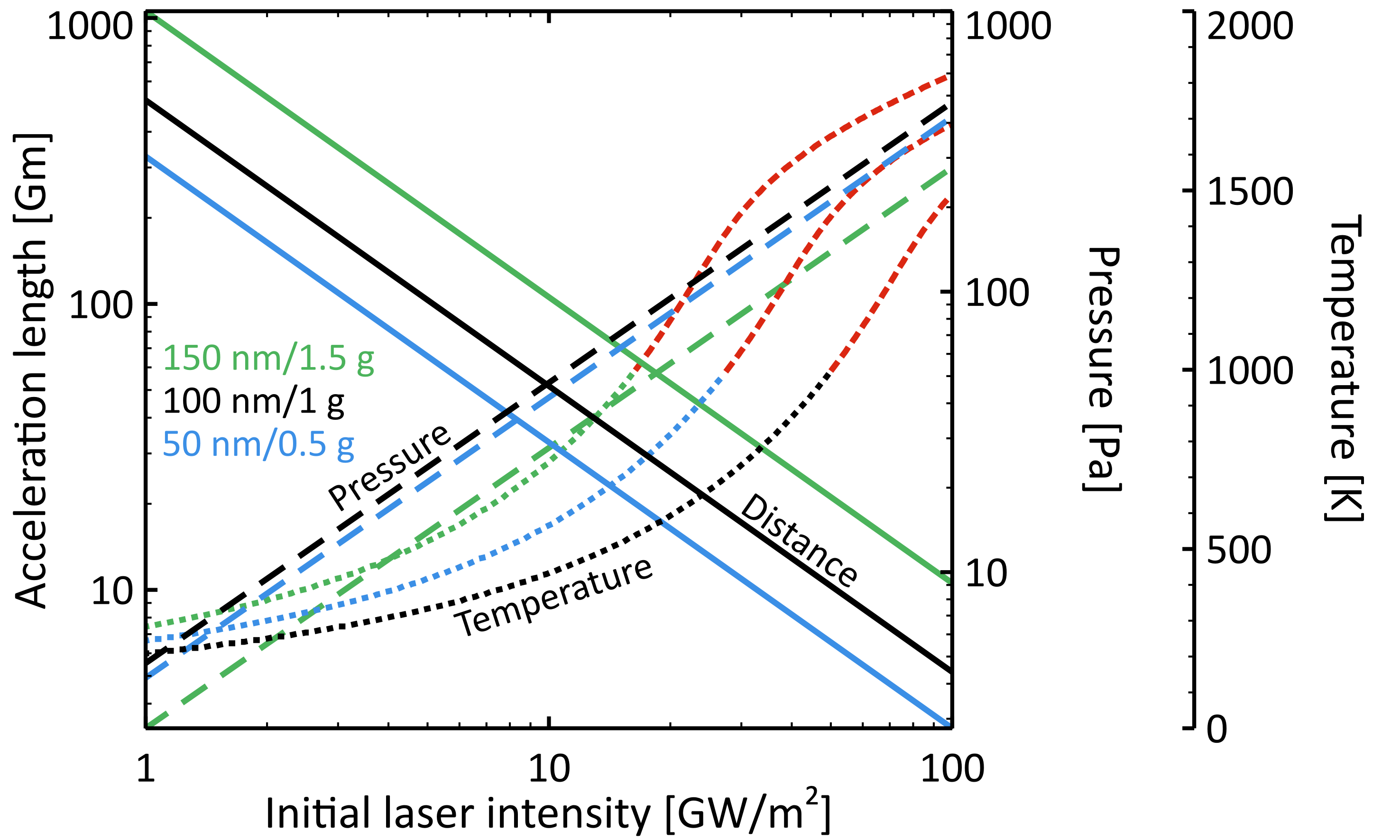}%
\caption{Acceleration length $L$ (solid lines), peak center pressure $P_{max}$ (dashed lines), and peak center temperature $T_{max}$ (dotted lines) of a flat (non-curved) $A_\perp=2$-\si{\meter\squared} area light sail composed of the simple \ch{MoS2}/\ch{Al2O3} hybrid material as a function of the initial laser light intensity on the sail $I_0=\frac{\Phi_0}{A_\perp}$ when the final velocity is $v_f=\frac{1}{5}c$ and the sail mass is $m_s=\frac{1}{2}m_{tot}$. Colors: green (sail film thickness~$t_f=150$~\si{\nano\meter}, sail mass~$m_s=1.5$~\si{\gram}), black ($t_f=100$~\si{\nano\meter}, $m_s=1$~\si{\gram}), and blue ($t_f=50$~\si{\nano\meter}, $m_s=0.5$~\si{\gram}). The red dashed lines indicate temperatures above the ultra-high vacuum sublimation point of \ch{MoS2} ($T_{sub}\approx1000$~\si{\kelvin})~\cite{Cannon1959-1612, Cui2018-44}. }%
\label{F:impactLaserPowerArea}%
\end{figure}

Before delving into the mechanical stresses of a curved sail, we first consider just the light pressure and the equilibrium temperature in the simple case of a flat (non-curved) light sail. We have accomplished this by way of Figure~\ref{F:impactLaserPowerArea}, in which we have adopted a final sail velocity of one fifth of the speed of light $\left(v_f=\frac{1}{5}c\right)$, used a laser output wavelength of $\lambda_0=1.2$~\si{\micro\meter}, and assumed that the sail mass is equal to the mass of the chip payload and tethers $\left(m_s=\frac{1}{2}m_{tot}\right)$, which is the optimal ratio for flat sails~\cite{Kulkarni2016-43}. Since $P$ and $T$ change throughout the acceleration phase (see \supinf), we have selected only the maximum values within each trajectory, $P_{max}$ and $T_{max}$, respectively, for this graphic. Importantly, cases with the lowest acceleration lengths $L$ suffer from excessively high pressures and temperatures, implying that mechanical and thermal factors place strong constraints on light sail designs and must be balanced against metrics like the acceleration length. 

Next, we consider the stress within the sail film that is caused by the photon pressure, for the case of a spherically-curved round sail that is illuminated by a non-polarized laser beam with a uniform (``top hat'') intensity profile. Though actual light sails may be parabolically curved and be accelerated using partially-polarized laser beams with Gaussian or Goubau intensity profiles~\cite{Parkin2018-370}, general design trade-offs can be obtained from this simpler case of uniform intensity. In the following, we will focus on the peak tensile stresses of the sail membrane, which generally occur at the sail's radial center, ignoring  local stresses such as those associated with the tether attachment points. 

Importantly, the tensile stress $\sigma$ experienced by the sail depends on its curvature.  In a flat (non-curved) sail, the light pressure is resisted by the the sail's bending stiffness, resulting in a maximum tensile stress of $\sigma_{flat}\sim\frac{P d_s^2}{t_f^2}$. In contrast, that of spherically curved sails is resisted by tensile membrane stretching and scales as $\sigma_{curved}\sim\frac{P s_s}{t_f}$~\cite{Timoshenko1959-book}. These relations suggest that the curved sail's stress is minimized for low spherical radii of curvature $s_s$ (\ie, very curved sails); however, geometrically, the minimum $s_s$ for a curved sail is on the order of the sail diameter $d_s$~\cite{Popova2016-1346}. Using the best case of $d_s \sim s_s$, the ratio of the stress in flat sails to that in curved sails can be estimated as $\frac{\sigma_{flat}}{\sigma_{curved}} \sim \frac{d_s}{t_f} \gg 1$ because $t_f(\sim100~\si{\nano\meter}) \ll d_s(\sim1~\si{\meter})$. This means that curved sails experience much lower stresses than flat sails; consequently, curved sails can sustain higher photon pressures $P$, allowing them to achieve lower acceleration lengths $L$ relative to flat sails. 

\begin{figure*}[ht!]
\centering
\includegraphics[width=\figWidthFull]{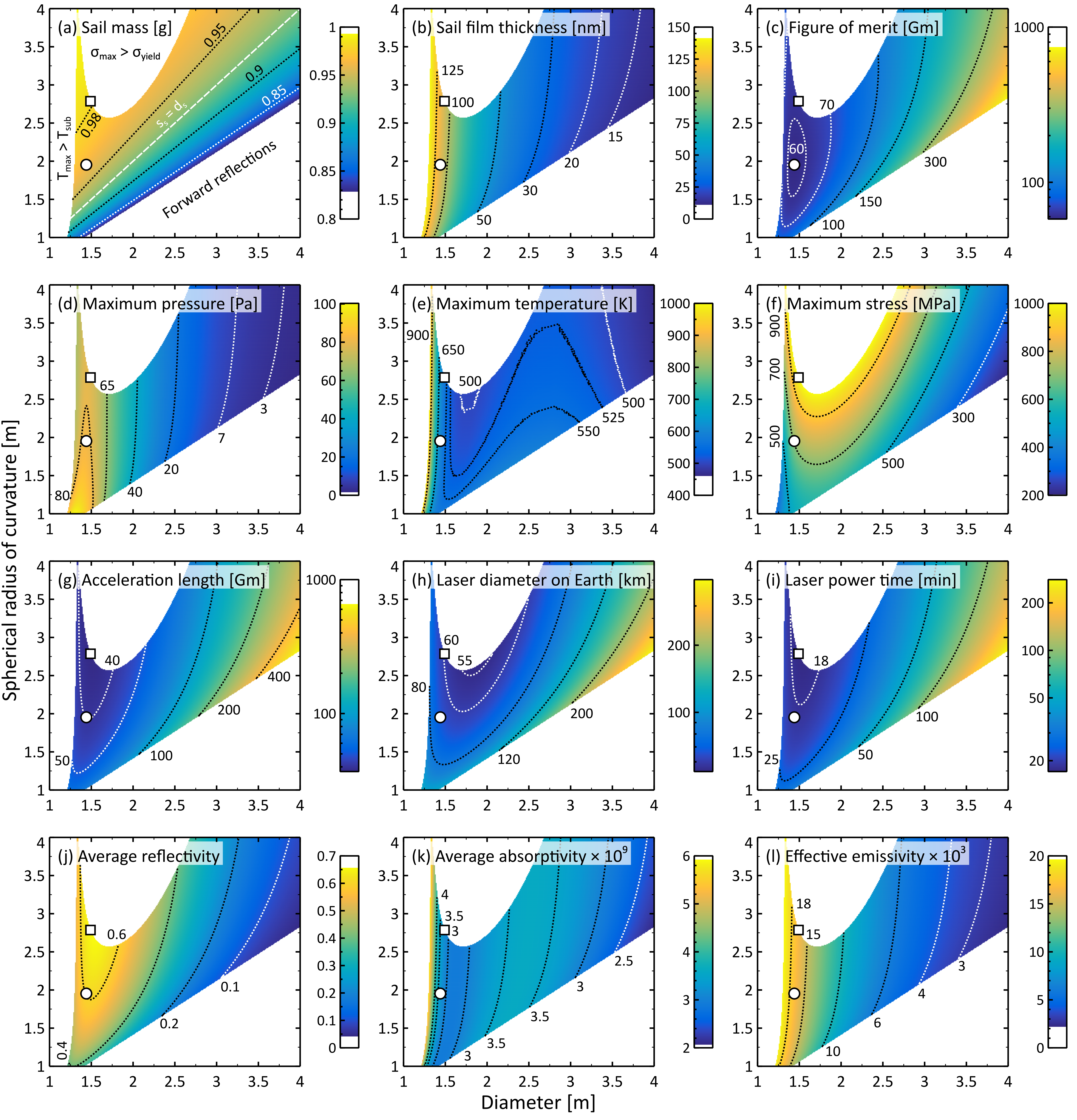}%
\caption{(a) Sail mass $m_s$, (b) film thickness $t_f$, (c) figure of merit $\fom$, (d) maximum pressure $P_{max}$, (e) maximum temperature $T_{max}$, (f) maximum stress $\sigma_{max}$, (g) acceleration length $L$, (h) laser diameter on Earth $d_{l,E}$, (i) laser power time $t_l$, (j) average reflectivity $\varrho_a$, (k) average absorptivity $\alpha_a$, and (l) effective emissivity $\varepsilon_e$ for a light sail composed of the simple \ch{MoS2}/\ch{Al2O3} hybrid material with $m_{tot}=2$~\si{\gram}, $\Phi_0=30$~\si{\giga\watt}, and $v_f=\frac{1}{5}c$. The dashed line in panel (a) denotes circular diameter and spherical radius of curvature equivalence: $d_s = s_s$. The squares denote the $d_s$-$s_s$ design exhibiting the minimum value of $L$, and the circles denote that with the minimum value of $\fom$. The $\alpha_a$ values of panel (k), though small, reflect the low extinction coefficients $\kappa$ of \ch{MoS2} and \ch{Al2O3} near the laser wavelength~\cite{Ermolaev2020-21, Lingart1982-706, Querry1985-report} and are comparable to values used elsewhere~\cite{Parkin2018-370}. }%
\label{F:diaRadSingle}%
\end{figure*}

We now explore the impact of a sail's diameter $d_s$ and radius of curvature $s_s$ on its performance in order to provide general guidelines for light sail design. To aid this discussion, we have plotted twelve relevant quantities, including $m_s$, $t_f$, $P_{max}$, $T_{max}$, $\sigma_{max}$, $L$, $\varrho_a$, $\alpha_a$, $\varepsilon_e$, and three others that we will introduce below, as a function of $d_s$ and $s_s$ in Figure~\ref{F:diaRadSingle} for a $m_{tot}=2$~\si{\gram} sailcraft accelerated by a constant laser output power of $\Phi_0=30$~\si{\giga\watt} (see \supinf). In these panels, as labeled in Figure~\ref{F:diaRadSingle}(a), we have removed designs for which the curvature would result in multiple reflections and cause excessive heating in the center of the sail $\left( s_s<\frac{d_s}{\sqrt{2}} \right)$, designs hot enough to sublimate the material, $T_{max}>T_{sub}\approx1000$~\si{\kelvin}~\cite{Cannon1959-1612, Cui2018-44}, and designs that would tear, $\sigma_{max}>\sigma_{yield}\approx1000$~\si{\mega\pascal}~\cite{Graczykowski2017-7647, Sledzinska2020-1169}. Importantly at low diameters $d_s$, few persisting light sail designs exhibit large radii of curvature $s_s$, which underscores the the imperative that light sails should be curved (higher $s_s$ values imply flatter structures). Put another way, light sails must be specifically designed to bow or billow in order to reduce the peak stresses they experience~\cite{Heinrich1966-52, Gordon1978-book, Sakamoto2007-514}. 

Further examination of Figure~\ref{F:diaRadSingle} reveals that most of the persisting designs feature similar diameter and radius of curvature values. This is highlighted by the white dashed line in panel (a), for which $d_s=s_s$, demonstrating a general principle that the diameter and radius of curvature of light sails must increase together. This fact can be explained by the scaling relationships introduced earlier, as well as by panels (b) and (d). Sails that have larger diameters (and hence larger perpendicular areas, $A_\perp=\frac{\pi}{4}d_s^2$) experience lower pressures, and sails that have lower radii of curvature (sails that are more curved) experience lower tensile stress and strain ($\epsilon\sim\frac{\sigma}{E}\times100\%$); such sails are therefore less likely to tear. It is worth noting that, at the $d_s=s_s$ condition, the surface area of a spherically curved sail $A_s$ is only about 7\% greater than that of a flat circular sail $A_\perp$, implying that appropriately-designed curved sails have comparable masses and thicknesses to flat sails. 

Panels (a) and (b) show the sail mass $m_s$ and film thickness $t_f$ values used in these calculations, respectively. In flat round sails, the optimum sail mass is half the total mass~\cite{Kulkarni2016-43, Ilic2018-5583}; however, the optimal mass of spherically curved round sails depends on $d_d$ and $s_s$, and specifically, $m_s$ decreases with increasing sail curvature (decreasing $s_s$; see panel (a) and \supinf). Importantly, this implies that, as the curvature of a sail increases, it is able to efficiently tow larger chip payloads within the same total mass budget. Next, notice from panel (b) that, due to the $m_{tot}=2$ \si{\gram} constraint, the sail can be only tens of nanometers thick,  emphasizing the importance of curvature to relieve stress.

The calculated acceleration length $L$ values are provided in panel (g), and the $d_s$-$s_s$ design corresponding to the minimum $L$ is indicated with the white square. Comparing with panel (f), however, shows that this point corresponds to $\sigma_{max}=\sigma_{yield}$, which has no failure margin. To identify designs with a higher probability of survival, we introduce a figure of merit $\fom$ 
\begin{equation}
    \fom = L \sqrt{1 + \left(\frac{\sigma_{max}}{\sigma_{yield}}\right)^2 + \left(\frac{\epsilon_{max}}{\epsilon_{crit}}\right)^2 + \left(\frac{T_{max}}{T_{sub}}\right)^2}
\label{E:fomBasic}
\end{equation}
that accounts for the mechanical and thermal limitations of the sail ($\epsilon_{crit}\approx5\%$~\cite{Graczykowski2017-7647, Sledzinska2020-1169}). Values of $\fom$ are provided in panel (c), and the design corresponding to the minimum $\fom$ is shown with the white circle. This sail features a similar $d_s$ but a lower $s_s$ compared to the minimum $L$ design, reducing the maximum tensile stress and strain by 29\% while increasing the acceleration length by only 8\%. The minimum-$\fom$ design also tows a 2\% heavier chip-tether payload. This simplified figure of merit underscores the importance of including both mechanical and thermal factors when optimizing light sail designs. 

The reason that the acceleration length $L$ is often used as a figure of merit for light sail design is that it determines the diffraction-limited diameter of the phased laser array on Earth $d_{l,E}\sim\lambda_0\frac{L}{d_s}$ and the time duration for which the array must produce light $t_l\sim \frac{L}{v_f}$, which are important engineering design criteria for the photon engine~\cite{Kulkarni2016-43, Kipping2017-277, Kulkarni2018-155, Kipping2018-103, Ilic2018-5583, Parkin2018-370}. These quantities are provided in panels (h) and (i). Notice that the $L$ and $t_l$ minima occur at similar $d_s$ values, whereas the minimum $d_{l,E}$ occurs at slightly larger $d_s$, reflecting the laser array diameter's inverse sail diameter scaling. Curvature allows light sails to be designed with smaller diameters, allowing them to achieve lower acceleration lengths and to be powered by smaller Earth-bound laser arrays for shorter laser power times. 

Lastly, we examine the impact of the sail diameter and radius of curvature on the optical properties of the sail, and the corresponding implications for the light sail's performance. To begin, we stress that the optical properties of the sail film vary greatly with its thickness and also depend on its photonically-engineered structure~\cite{Ilic2018-5583, Myilswamy2020-8223, Salary2020-1900311, Weiliang2020-2350}. The film thickness in our calculations depends on the sail's diameter and spherical radius of curvature, since in Figure~\ref{F:diaRadSingle} we hold the total sailcraft mass $m_{tot}$ constant (see panel (b) and the \supinf). The $d_s$ dependence is the cause of the horizontal variations in optical property values observable in panels (j)-(l), which show that $\varrho_a$, $\alpha_a$, and $\varepsilon_e$ generally increase with increasing $t_f$ (\ie, decreasing $d_s$). Importantly, however, panels (j) and (k) reveal local reflectivity maxima and absorptivity minima, respectively, near $d_s\approx1.5$~\si{\meter} ($t_f\approx100$~\si{\nano\meter}), which are related to interference effects due to phases acquired by the electromagnetic wave as it travels in the film~\cite{Macleod2017-book}. The film thickness corresponding to such optimal reflection scales inversely with the refractive index of the material, $t_f \sim \frac{\lambda_0}{4n}$, but is more accurately discovered using a time-dependent numerical optimization that takes into account the laser light's Doppler shift due to the increasing light sail velocity. In contrast to the reflection and absorption, the emissivity values in panel (l) do not show any local maxima or minima, in part because $\varepsilon_e$ is calculated over a large wavelength range, thereby smoothing over local variations, whereas $\varrho_a$ and $\alpha_a$ are determined over a much smaller interval corresponding to the laser wavelength and its Doppler-shifted range. The combined impacts of changes in $\alpha_a$ and $\varepsilon_e$ can be most clearly seen in the $T_{max}$ values of panel (e). In particular, at low diameters, as $d_s$ decreases, the ratio $\frac{\alpha_a}{\varepsilon_e}$ increases, leading to higher sail temperatures. 

In addition, we may observe that $\varrho_a$ (panel (j)) has a strong positive dependence on $s_s$. This arises primarily because flatter sails (larger $s_s$) produce a larger reflected component in the longitudinal direction (back toward the laser) and because off-perpendicular polarity-averaged reflectivity values, occurring at the perimeter of a curved sail, are generally lower than normal-reflection values. The impacts of this radius-dependent reflectivity can be most clearly seen in the acceleration length $L$ values of panel (g), which reveal that, for limited laser output power, flatter sails achieve lower acceleration lengths (in Figure~\ref{F:diaRadSingle}, $\Phi_0=30$~\si{\giga\watt} is held constant for all $d_s$-$s_s$ sail designs). Finally, we note that the maximum photon pressure $P_{max}$ contour shapes in panel (c) differ from those of the average reflectivity $\varrho_a$ (panel (j)) because $P_{max}$ depends on $\varrho_\perp$ (not shown in Figure~\ref{F:diaRadSingle}) rather than on $\varrho_a$ (see \supinf). 

\begin{figure*}[ht!]
\centering
\includegraphics[width=130mm]{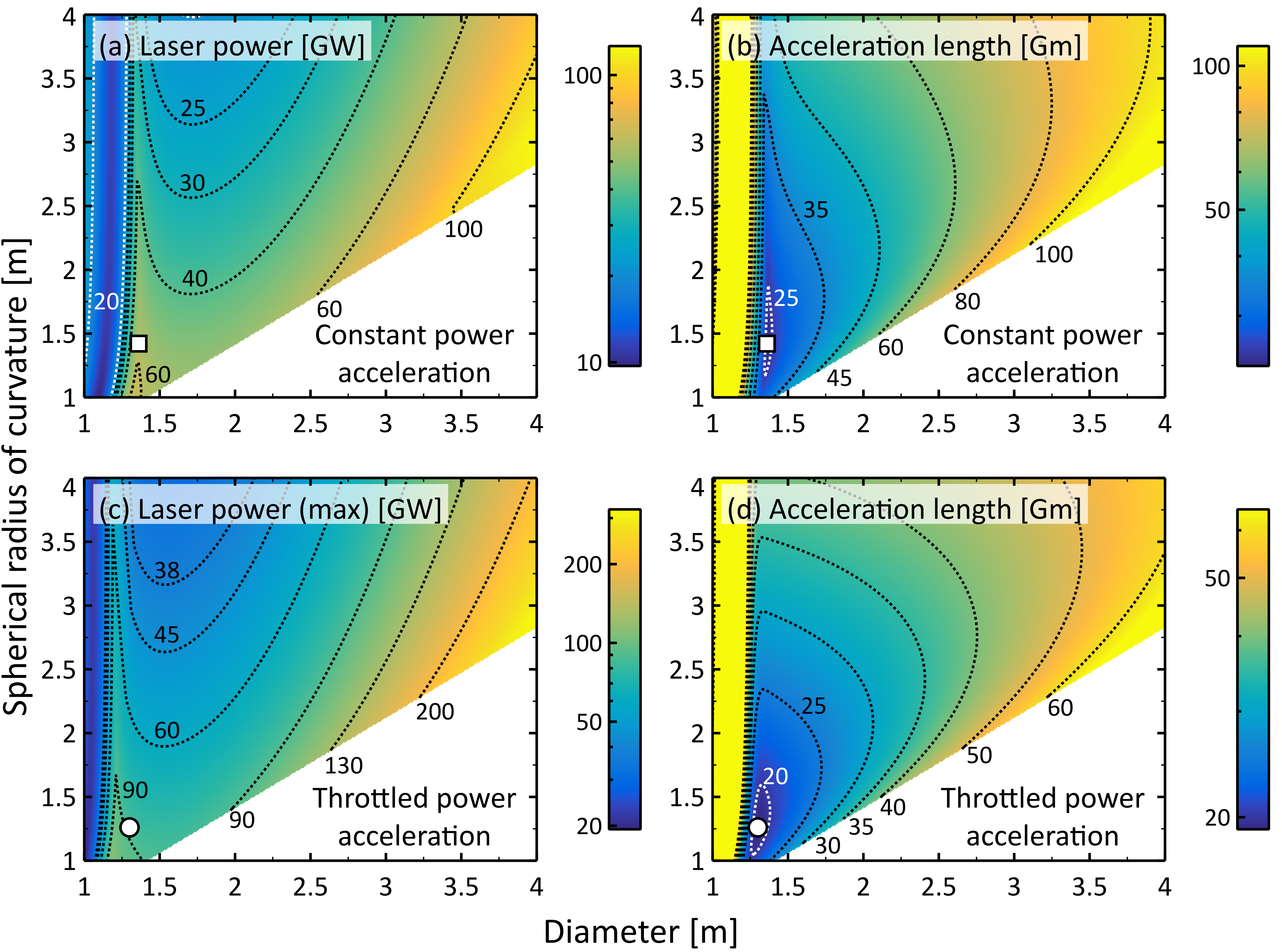}%
\caption{Laser power $\Phi$ and acceleration length $L$ for a $m_{tot}=2$~\si{\gram} starcraft composed of the simple \ch{MoS2}/\ch{Al2O3} hybrid material using constant laser output power acceleration (top row) and throttled laser output power acceleration (bottom row)~\cite{Parkin2018-370}, with $v_f=\frac{1}{5}c$ and $\xi=0.001$. The squares and circles denote the $d_s$-$s_s$ designs exhibiting the minimum values of $L$ for constant laser output power and throttled laser output power acceleration, respectively. The designs shown in this figure have the same $m_s$,  $t_f$, $\varrho_a$, $\alpha_a$, and $\varepsilon_e$ values as those displayed in Figure~\ref{F:diaRadSingle} panels (a), (b), (j), (k), and (l), respectively. }%
\label{F:diaRadThrottlingSingle}%
\end{figure*}

What remains is to demonstrate that, if sufficient laser output power is available, curved light sails will achieve shorter acceleration distances $L$ than their flatter counterparts.  We have accomplished this in Figure~\ref{F:diaRadThrottlingSingle}, which reflects two laser power modulation approaches. In the top row, for each $d_s$-$s_s$ design, we iteratively determined the constant laser output power $\Phi_0$ for which the greatest of $\frac{\sigma_{max}}{\sigma_{yield}}$, $\frac{\epsilon_{max}}{\epsilon_{crit}}$, and $\frac{T_{max}}{T_{sub}}$ was equal to $1-\xi$, \ie, $\text{max}\big\{\frac{\sigma_{max}}{\sigma_{yield}} \text{,} \frac{\epsilon_{max}}{\epsilon_{crit}} \text{,} \frac{T_{max}}{T_{sub}}\big\} = 1-\xi$, where $\xi=0.001$ is the failure margin. Thus, this first strategy, which we will call constant laser output power acceleration, ensures that no light sail designs are exercised beyond their failure points, and that all designs are operated at a fraction $1-\xi$ of their failure points for at least one instant during their acceleration phases. In the bottom row, for each $d_s$-$s_s$ design, at each fractional speed of light value $\beta=\frac{v}{c}$ of the acceleration phase ($0<\beta<0.2$), we iteratively determined the laser output power $\Phi_\beta$ for which $\text{max}\big\{\frac{\sigma_\beta}{\sigma_{yield}} \text{,} \frac{\epsilon_\beta}{\epsilon_{crit}} \text{,} \frac{T_\beta}{T_{sub}}\big\} = 1-\xi$. This second approach, referred to as throttled laser output power acceleration~\cite{Parkin2018-370}, thereby ensures that all light sail designs are operated at a fraction $1-\xi$ of their failure points throughout the entire durations of their acceleration phases (see \supinf). Laser power throttling can be accomplished, for instance, by activating laser modules on the diameter of the phased array as greater power is required in the acceleration phase~\cite{McManamon2005-152, Parkin2018-370, Leger2020-65, Bandutunga2021-1477}. This has the added benefit of increasing the diffration-limited distance over which the laser can be focused on the light sail~\cite{Ilic2018-5583}. Panels (a) and (c) show the $\Phi_0$ values and the $\Phi_{max}$ results, respectively, and panels (b) and (d) show the calculated $L$ values. The minimum $L$ values are seen to occur in sails with small diameters and significant curvature (small $s_s$), reflecting our earlier observation that curvature decreases the stress and strain felt by light sails and allows them to endure greater laser power levels. Also, though not shown in Figure~\ref{F:diaRadThrottlingSingle}, we note that the laser array diameters on Earth $d_{l,E}$ and the laser power times $t_l$ are also minimized for sails with significant curvature. Although these calculations were performed with a minimal failure margin of $\xi=0.001$, similar trends are observable for more reasonable failure margins such as $\xi=0.1$ (see \supinf). In addition to the lower $L$ values, curved sails also accommodate larger chip payloads; the minimum $L$ design in panel (d) tows a chip-plus-tether mass that is 8\% greater than that of a comparable flat round sail. Photonic engineering of the sail could further reduce the film thickness required to achieve adequate optical properties, enhancing these results~\cite{Ilic2018-5583, Myilswamy2020-8223, Salary2020-1900311, Weiliang2020-2350}. Finally, we mention that, since beam-riding stability requirements dictate that the center of mass of the sail-tether-chip system must be further from the sail than the sail's focal point $f\sim\frac{s_s}{2}$, low $s_s$ values have an additional benefit in that they allow for shorter sail-chip tether lengths and hence lower tether masses~\cite{Popova2016-1346}.

To summarize, we have observed that photon pressures and sail temperatures increase inversely with the acceleration length, such that mechanical factors constrain feasible light sail designs. A convenient method to alleviate excessive stress and strain is to allow light sails to curve, \ie, to bow or billow, as they accelerate, and a general rule of thumb is that the light sail's spherical radius of curvature should be comparable to its diameter. We have noted the presence of optimal sail film thicknesses that exhibit high reflectivity and low absorptivity near the laser wavelength, and high emissivity at longer thermal emission wavelengths. We also observed that the sail's reflectivity decreases as it becomes more curved (\ie, as its spherical radius of curvature decreases), suggesting a trade-off between the sail's mechanical integrity and its optical attributes. Additionally, we used a figure of merit to suggest designs that achieve low acceleration lengths and are also mechanically and thermally robust. Finally, we demonstrated that, given sufficient laser output power, increasing a light sail's curvature allows it to achieve lower acceleration distances and to tow larger payload chips. These results will serve as guides for future light sail designs, enabling heretofore impossible investigations of deep space, including the Alpha Centauri system. 

\section*{Acknowledgements}\label{S:ack}

This work was supported by the Breakthrough Initiatives, a division of the Breakthrough Prize Foundation. It was also funded in part by a National Science Foundation CAREER award under grant CBET-1845933. J.B.\ is supported by a National Science Foundation Graduate Research Fellowship under grants DGE-1650605 and DGE-2034835. The authors thank Prof.\ Prashant K.\ Purohit for useful discussions about thin film mechanics, as well as Mohsen Azadi, Thomas J.\ Celenza, Dr.\ Pawan Kumar, Jason Lynch, and George A.\ Popov for insightful comments. 

%% bibdatabase file
\small
\bibliography{sailMechSources}

\end{document}